\documentclass[final,12pt,onecolumn]{IEEEtran}
\usepackage{amsmath,amsfonts,cite}

\long\def\comment#1{}

\newcounter{assumption}

\long\def\comment#1{}
\newtheorem{theorem}{Theorem}

\begin{document}

\linespread{1.55}\normalsize \tolerance 5000

\title{Comments on Degrees of freedom region for  $K$-user interference channel with $M$ antennas }

\author{ \normalsize
 \IEEEauthorblockN{Huarui Yin}\\
 \small
 \IEEEauthorblockA{WINLAB, Department of Elec.\ Eng.\ and
 Info.\ Sci.\\ University of Sci.\ and Tech.\ of China, Hefei,
 Anhui, 230027, P.R. China\\ Email: yhr@ustc.edu.cn
 }
}
\maketitle

\begin{abstract}

For a $K$-user interference channel with $M$ antenna at each transmitter and
each receiver, the maximum total DoF of this channel has been previously
determined to be $\max \sum_{k=1}^K d_k = MK/2$. However, the DoF region
remains to be unknown. In this short note, through a simple time-sharing
argument, we obtain the degrees of freedom (DoF) region of this channel.

\end{abstract}

Consider a $K$-user interference channel with $M$ antennas at each transmitter
and each receiver, the same as in \cite{vj08}. Let $d_k$ denote the degrees of
freedom (DoF) of user $k$, $k=1,\ldots, K$. The maximum total DoF
$\max\sum_{k=1}^K$ has been found in \cite{vj08} to be $MK/2$. We have the
following result regarding the DoF region.

\begin{theorem}
The degrees of freedom region of the $K$-user interference channel with
$M$-antennas at both receivers and transmitters is characterized as follows :
\begin{equation}\label{eq.dof.region}
\mathcal{D} = \left\{ \left( {{d_1},{d_2}, \cdots ,{d_K}} \right):
\quad {d_i} + {d_j} \le M, \forall 1 \le i,j \le K,i \ne j \right\}.
\end{equation}
\end{theorem}
\begin{IEEEproof}
The converse argument is the same as the converse argument in
\cite[Theorem~1]{vj08}. We show the achievability as follows.

Without loss generality, suppose $d_1^*\geq d_2^* \geq d_k^*,k = 3,\cdots, K$,
and $d_i^*+d_j^*\le d_1^*+d_2^*\le M$, $\forall i,j\in [1, K]$. We would like
to show that $(d_1, d_2, \ldots, d_M)=(d_1^*, d_2^*, \ldots, d_M^*)$ is
achievable.

It is obvious that
\[
    (d_1, d_2, \ldots, d_K)=(M,0,\ldots, 0)
\]
can be achieved by single user transmission. It is also known from \cite{vj08}
that the point
\[
    (d_1, d_2, \ldots, d_K)=(M/2, M/2,\ldots, M/2)
\]
is achievable. Trivially, the point
\[
    (d_1, d_2, \ldots, d_K)=(0,0, \ldots, 0)
\]
is achievable.

By time sharing, with weights $(d_1-d_2)/M$, $2d_2/M$ and
$1-d_1/M-d_2/M$ among the three points, in that order, it follows
that the point
\[
(d_1, d_2, \ldots, d_K)=(d_1^*, d_2^*, d_2^*, \ldots, d_2^*)
\]
is achievable. This is already no smaller than the DoF we would like to have.

\end{IEEEproof}

\bibliography{refs}
\bibliographystyle{IEEE}
\end{document}